\documentclass[prl,amsmath,aps,twocolumn,superscriptaddress]{revtex4}
\usepackage{amsfonts}
\usepackage{graphicx,multirow,color}
\begin{document}
\title{Tripartite entangled pure states are tripartite nonlocal}

\author{Sixia Yu}
\affiliation{Centre for quantum technologies, National University of Singapore, 3 Science Drive 2, Singapore 117543, Singapore}
\author{C.H. Oh}
\affiliation{Centre for quantum technologies, National University of Singapore, 3 Science Drive 2, Singapore 117543, Singapore}
\affiliation{Physics department, National University of Singapore, 3 Science Drive 2, Singapore 117543, Singapore}

\begin{abstract}
Nonlocal correlations as revealed by the violations to Bell inequalities are incompatible with local models without any nonlocal correlations. However some tripartite entangled states, e.g.,  symmetric pure states, exhibit a stronger nonlocality even incompatible with hybrid local/nonlocal models allowing nonlocal correlations between any two parties. Here we propose a three-particle Hardy-type test without inequality, as illustrated by a gedanken experiment, that is failed by all non-signaling local models while is passed by all tripartite entangled {\em asymmetric} pure states. Our result implies that every tripartite entangled state, regardless of the dimensions of the underlying Hilbert spaces, passes a Hardy-type test and therefore is genuine tripartite nonlocal. Maximal success probability and a generalization to multipartite cases of our test are also presented.

\end{abstract}


\maketitle

Entanglement and nonlocality are two different but intrinsically related quantum features, both playing important roles in our fundamental understandings of  physical world as well as in various novel quantum informational tasks. It was first shown by Bell \cite{bell} that there are entangled states that do not admit any local realistic models and therefore are nonlocal. Much later Werner \cite{werner}, after giving an accurate definition to entangled mixed states,  showed that there are entangled states that admit local realistic models. For pure states, however, Gisin \cite{gisin} proved that  all entangled bipartite pure states violate a Bell inequality \cite{chsh}, and therefore are nonlocal. Gisin's result was generalized to multipartite scenario \cite{pr,chen,3,yu} and specially it was shown \cite{yu} that all the pure states violate a single Bell inequality arising from Hardy's test for nonlocality without inequality.

The local realistic models excluded by Bell inequalities, as well as other ingenious approaches such as GHZ paradoxes \cite{ghz,mermin} or Hardy's test without inequality \cite{hardy}, forbid any sorts of nonlocal correlations. In the case of three particles, however, nonlocality can still be involved in a `local' model, e.g., any two particles out of three may share some nonlocal correlations. Imagine that three particles are distributed among three distant observers $1,2$, and $3$, who may perform some two-outcome measurements. In a most general hybrid local/non-local model, the joint probability of three observers measuring observables $a_1,a_2$, and $a_3$  with outcomes, e.g., $0,1$, and $0$, respectively, has a bi-local form
\begin{eqnarray}
\label{bl}
P(a_1\bar a_2a_3)=\int d\lambda\, \varrho_\lambda\label{bl1}
L_\lambda(a_1)S_\lambda(\bar a_2a_3)+\hskip 1cm\cr \int d\lambda\, \varrho_\lambda^\prime
L^\prime_\lambda(\bar a_2)S^\prime_\lambda(a_1a_3)+\hskip 0.5cm \cr\int d\lambda\, \varrho_\lambda^{\prime\prime}
L^{\prime\prime}_\lambda(a_3)S^{\prime\prime}_\lambda(a_1\bar a_2).\label{bl3}
\end{eqnarray}
Here $S_\lambda^{(\prime,\prime\prime)}$ and $L_\lambda^{(\prime,\prime\prime)}$ are some bipartite and single-particle probability distributions, respectively,  for a given hidden variable $\lambda$ distributed according to $\varrho_\lambda^{(\prime,\prime\prime)}$, respectively. It was first shown by Svetlinchy \cite{svet} that, via the violation of a Bell-type inequality, some tripartite entangled states even do not admit such a general kind of local models and therefore these states exhibit genuine tripartite nonlocality. Svetlinchy's result has also been generalized to different cases \cite{sveta1,sveta2,svetb1,svetb2}.

However, Svetlincy's notion of genuine multipartite nonlocality is so general that nonlocal correlations shared by part of observers may even allow signaling. Moreover it is found numerically that certain tripartite entangled pure state does not violate any Svetlinchy-like inequalities  \cite{def}. Since a reasonable physical theory should not permit signaling, it is natural to work only with  non-signaling correlations. In particular, all bipartite joint probability distributions appeared in Eq.(\ref{bl}), should satisfy non-signaling constraints such as
\begin{subequations}\label{ns}
\begin{eqnarray}
S_\lambda(a_2b_3)+S_\lambda(\bar a_2 b_3)&=&S_\lambda(b_2b_3)+S_\lambda(\bar b_2 b_3),\label{ns1}\\
S_\lambda(a_2b_3)+S_\lambda(a_2 \bar b_3)&=&S_\lambda(a_2a_3)+S_\lambda(a_2 \bar a_3),\label{ns2}
\end{eqnarray}
\end{subequations}
where $a_{2,3},b_{2,3}$ are two different measurement settings for particles 2 and 3. A genuine multipartite correlation should also exclude such kinds of non-signaling  local models \cite{def}.  One fundamental task now is to ascertain whether or not a given set of joint probabilities can be casted into the bi-local form Eq.(\ref{bl}).

Besides the statistical approach via Bell-type inequalities \cite{svet,def}, Hardy-type tests without inequality have also been proposed recently. One Hardy-type argument \cite{rwz} is introduced to detect genuine multipartite entanglement, singling out a special kind of entangled pure states. Another Hardy-type argument \cite{chenq} is introduced to detect genuine multipartite nonlocality and all entangled symmetric pure states are found to be genuine multipartite nonlocal. Though there are strong numerical evidences for the violations of a Bell-type inequality \cite{def} and the survival of a Hardy-type test \cite{chenq}  by all fully entangled 3-qubit pure states, it is still unknown even for tripartite systems whether these two notions, genuine multipartite entanglement and genuine multipartite nonlocality, are equivalent or not for pure states.

In this Letter we completely solve this problem for three particles with the help of another Hardy-type test for genuine tripartite nonlocality without inequality. After showing that our test is failed by all non-signaling local models, we propose a gedanken experiment to illustrate a Hardy-type paradox for genuine tripartite nonlocality. Then we identify all the fully entangled 3-qubit pure states that pass or fail our test. Via a local projection to a 3-qubit subspace, every tripartite entangled pure state, regardless of the dimensions of underlying Hilbert space, also passes a Hardy-type test and thus is genuine tripartite nonlocal. A generalization  to multipartite cases and the maximal survival probability of our test are also presented.

Suppose that each observer $k=1,2,3$ measures two dichotomic observables $a_k$ and $b_k$ with outcomes labeled by $0$ and $1$. We propose the following six conditions
\begin{subequations}\label{ht}
\begin{eqnarray}
P(a_1a_2a_3)&>&0,\label{ht1}\\
P(a_1a_2\bar b_3)&=&0,\label{ht4}\\
P(a_1\bar b_2 a_3)&=&0,\label{ht5}\\
P(a_1b_2 b_3)&=&0,\label{ht2}\\
P(b_1a_2 b_3)&=&0,\label{ht3}\\
P(\bar b_1\bar b_2 b_3)&=&0.\label{ht6}
\end{eqnarray}
\end{subequations}
to detect genuine tripartite nonlocality. We observe that, firstly, if the last condition is replaced with the condition $P(\bar b_1a_2a_3)=0$ then the Hardy-type test proposed in \cite{chenq} is reproduced in the case of three particles. Secondly, the first four conditions in the above test are exactly Hardy's test for bipartite nonlocality without inequality for subsystem 2 and 3, conditioned on the first observer's measurement of $a_1$ with $0$ as outcome.

All non-signaling local models fail our Hardy-type test Eq.(\ref{ht}), in which every joint probability appearing in our test Eq.(\ref{ht}), e.g., $P(a_1a_2a_3)$, has exactly the same bi-local form as in Eq.(\ref{bl}) for $P(a_1\bar a_2a_3)$. In fact, because of the second observation made above the first four conditions cannot be simultaneously satisfied if particle 2 or 3 does not share nonlocal correlations with others. Suppose now particle 1 is local, i.e.,  there are some hidden variables $\lambda$ such that $L_\lambda(a_1)S_\lambda(a_2a_3)>0$ because of the first condition Eq.(\ref{ht1}). However this is impossible due to the following  contradiction
\begin{eqnarray}\label{cd}
0&=&L_\lambda (\bar b_1)S_\lambda( \bar b_2 b_3)+L_\lambda (a_1)S_\lambda (b_2  b_3)\nonumber\\
&\ge&\min \{L_\lambda (a_1),L_\lambda (\bar b_1)\}[S_\lambda(  b_2 b_3)+S_\lambda (\bar b_2  b_3)]\nonumber\\
&=&\min \{L_\lambda (a_1),L_\lambda (\bar b_1)\} [S_\lambda(a_2 b_3)+S_\lambda (\bar a_2 b_3)]\nonumber\\
&\ge& [L_\lambda (a_1)-L_\lambda(b_1)]S_\lambda(a_2 b_3)\nonumber\\&=& L_\lambda(a_1)S_\lambda (a_2 b_3)\nonumber\\&=&L_\lambda (a_1)[S_\lambda(a_2 b_3)+S_\lambda (a_2\bar b_3)]\nonumber\\
&=&L_\lambda (a_1)[S_\lambda(a_2\bar a_3)+S_\lambda(a_2 a_3)]>0.
\end{eqnarray}
Here the first equality is because of conditions Eq.(\ref{ht2}) and Eq.(\ref{ht6}) and the second and last equalities are due to the non-signaling constraints Eq.(\ref{ns1}) and Eq.(\ref{ns2}), respectively, while the third and fourth equalities are due to conditions Eq.(\ref{ht3}) and Eq.(\ref{ht4}), respectively. The second inequality is due to the fact that $L_\lambda(b_1)\ge 0$ and $L_\lambda(a_1)\le 1=L_\lambda(b_1)+L_\lambda(\bar b_1)$.  To conclude, conditions in Eq.(\ref{ht}) are incompatible with the bi-local form of joint probability distributions.  Thus if a tripartite quantum mechanical state passes our test, i.e., all conditions in Eq.(\ref{ht}) are satisfied, then the state cannot have a non-signaling local model and therefore is  genuine tripartite nonlocal. Let us elucidate our test by an example first.

\begin{figure}
\includegraphics[scale=0.9]{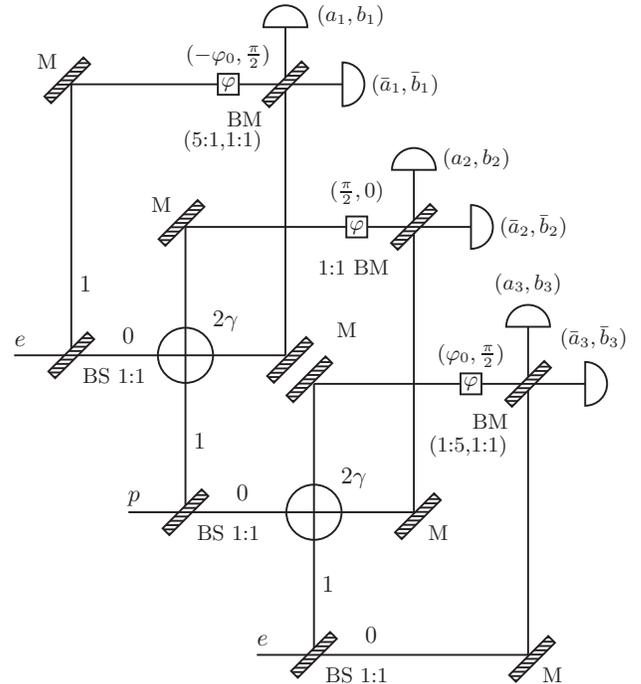}
\caption{A gedanken experiment for testing genuine tripartite nonlocality without inequality. Each observer has the freedom of choosing two alternative measurement settings characterized by the phase shifts and transmission-to-reflection ratios of the beam mergers. For example if observer 3 measures the first observable $a_3$ then the phase shifter $\varphi_0=\arcsin \frac1{\sqrt5}-\pi$ and 1:5 beam merger are chosen with upper and lower detectors corresponding to outcomes 0 and 1 respectively.}
\end{figure}

We consider a gedanken experiment similar to the one given by Hardy for two particles that includes three overlapped Mach-Zehnder interferometers for two electrons and a positron as depicted in Fig.1. When the first electron takes the lower path 0 and the positron takes the upper path 1, or the second electron take the upper path and the positron takes the lower path, then they annihilate and produce a photon pair.  
After the post-selection of the no-photon cases, the state of the whole system before the detectors reads
\begin{equation}
|000\rangle+|100\rangle+|110\rangle+|111\rangle.
\end{equation}
Consider the measurement settings $|b_1\rangle=|b_3\rangle=|a_2\rangle=|0\rangle+i|1\rangle$ and $|b_2\rangle=|0\rangle+|1\rangle$, $|a_1\rangle=|0\rangle+x|1\rangle$ and $|a_3\rangle=x|0\rangle+|1\rangle$ with $x=(i-2)/5$. It can readily be checked that all conditions in Eq.(\ref{ht}) are satisfied with  a small but nonzero success probability $P(a_1a_2a_3)=1/72$. For examples the detector $a_3$ never fires if both detectors $a_1$ and $\bar b_2$ are triggered as required by condition Eq.(\ref{ht4}) while it is possible for  $a_1,a_2,a_3$ to fire simultaneously, a fact incompatible with any non-signaling local models.

Suppose that the positron or the second electron is local while the other two particles may have some conspiracies, we have an effective 2-particle Hardy's paradox. Consider, e.g., the positron is local while two electrons may share some nonlocal correlations. The simultaneous clicks of detectors $a_1,a_2,a_3$ mean that i) $a_1$ never fires along with $\bar b_3$ due to Eq.(\ref{ht4}) and ii) $b_2$ always fires due to Eq.(\ref{ht5}) so that $a_1$ never fires along with $b_3$ due to Eq.(\ref{ht2}). As a result of non-signaling conditions $a_1$ never fires which contradicts the fact that $a_1$ fires along with $a_2$ and $a_3$. Suppose now that the first electron is local while the second electron and the positron have some conspiracies. A click of detector $a_1$ means that $b_3$ always fires along with $a_2$ and $b_3$ never fires along with $b_2$ due to Eq.(\ref{ht4}) and Eq.(\ref{ht2}), respectively. Due to Eq.(\ref{ht3}) the detector $b_1$ cannot fire so that $b_3$ is never triggered along with $\bar b_2$ due to condition Eq.(\ref{ht6}). As a result of non-signaling conditions, that $b_3$ is neither trigged along with $b_2$ nor with $\bar b_2$ infers that $b_3$ would not fire at all if observable $b_3$ were measured, which contradicts the fact that it would fire along with $a_2$.

Let us consider a general 3-qubit pure state $|\psi\rangle$ and in its magic basis, which is introduced in \cite{yu} to derive the violation of a single Bell inequality by all entangled pure states, we have expansion
\begin{equation}\label{3qs}
|\psi\rangle=h^*|000\rangle+u|011\rangle+v|101\rangle+s|110\rangle+t|111\rangle,
\end{equation}
with $|h|>0$. By choosing the phases of $|1\rangle_1$ and $|0\rangle_{1,2,3}$ we can always make $t,u,v,s\ge0$, respectively. The expansion above can be regarded as an alternative canonical form for 3-qubit pure state and, comparing with the one in \cite{acin}, has the advantage of being applicable to higher level systems. A magic basis of a given state can be specified by its closest product state (see Appendix).

A 3-qubit pure state $|\psi\rangle$ is symmetric if all three coefficients  $\{u,v,s\}:=A$ in its magic basis are equal and asymmetric otherwise. For an asymmetric state there are  at least two different elements in $A$ so that $|h|$ is different from at least one coefficient in $A$ and, by the freedom of choosing which  qubit to have label 3, we can always choose $s=0$ if $0\in A$ and $s\not=|h|$ if  $0\not\in A$. Apart from $s$ there is at least one nonzero element in $A$, and we can suppose $u>0$ by the freedom of choosing which qubit to have label 1. For a fully entangled $|\psi\rangle$ we have either $t>0$ or $t=0$ with at least two elements in $A$ being nonzero. In this case $s>0$ infers $u,v>0$ so that for a fully entangled asymmetric state we can always assume without loss of generosity
\begin{equation}\label{cdt}
|h|\not=s,\quad u>0, \quad t+s+v>0.
\end{equation}

{\it Theorem } The 3-particle Hardy-type test Eq.(\ref{ht}) is passed by all the fully entangled asymmetric pure states as well as by all the symmetric pure states with i) $t\not=0$, $s\not=0$, and $h\not=s$ or ii) $t=0$, $s\not=0$, and $|h|\not= s$, while is failed by all the other symmetric pure states.

{\it Outline of proof. }Detailed proof can be found in Appendix.   We denote by $|\varphi^*\rangle$ the complex conjugate state of an arbitrary single-qubit state $|\varphi\rangle$ in a given basis and $|\bar\varphi\rangle:=J|\varphi^*\rangle$ is orthogonal to $|\varphi\rangle$ with $J=i\sigma_y$. For later use we introduce three single-qubit operators $C=C_0+zC_1$, $\tilde C=z^*C_0-C_1$, and $D=xD_0+yD_1$ with  complex $x,y,z$ being arbitrary, where $C_{0,1}$ have matrix elements,  with $\mu,\nu,\tau=0,1$,
\begin{equation}
\langle\nu| C_\tau|\mu\rangle=\langle\psi|\mu,\nu, \tau\rangle,\quad \langle\mu| D_\tau|\nu\rangle=\langle\psi|\tau,\mu, \nu\rangle.
\end{equation}
In the magic basis of a given 3-qubit pure state $|\psi\rangle$, the last four conditions in  Eq.(\ref{ht}) determine the measurement settings to be,
\begin{eqnarray}\label{ms}
&|a_1\rangle=x|0\rangle+y|1\rangle,\quad |b_1\rangle=C^\dagger C|a_1\rangle;&\nonumber \\
&|a_2\rangle=J CC^\dagger C|a_1\rangle,\quad|b_2\rangle=J C|a_1\rangle;& \\
&|a_3\rangle=JD^T C^*|a_1^*\rangle,\quad |b_3\rangle=|0\rangle+z|1\rangle,&\nonumber
\end{eqnarray}
with $z$ being arbitrary and $(x,y)$ satisfying $|x|^2+|y|^2=1$ being determined by the second condition Eq.(\ref{ht4})
\begin{equation}\label{y}
\langle\psi |a_1 a_2 \bar b_3\rangle=\langle a_1^*|C^TC^* C^TJ\tilde C|a_1\rangle=0.
\end{equation}
For those states specified in the first part of Theorem, by excluding at most ten values of $z$, we can assure that the first condition Eq.(\ref{ht1}) is also satisfied. For those symmetric states specified by the second part of Theorem, i.e.,
 GHZ-like states $\cos\theta|000\rangle+\sin\theta|111\rangle$ and states
\begin{equation}
\frac12\sin\theta\big(|000\rangle+|011\rangle+|101\rangle+|110\rangle\big)+\cos\theta|111\rangle
\end{equation}
the solutions to Eq.(\ref{y}) always lead to $\langle\psi|a_1a_2a_3\rangle=0$ for arbitrary complex $z$, i.e., condition Eq(\ref{ht1}) cannot be satisfied so that these states fail our Hardy-type test. \hfill $\square$

Some remarks are in order. First, it has been shown in \cite{chenq} that all fully entangled symmetric pure 3-qubit states, especially those exceptional symmetric states that fail our test, pass another Hardy-type test for genuine tripartite nonlocality. Together with this result for symmetric cases, whose proof is included in Appendix for completeness, our result for asymmetric cases implies that every fully entangled 3-qubit pure state passes a Hardy-type test for genuine tripartite and therefore is genuine tripartite nonlocal. This shows a dramatical difference between the bipartite and tripartite nonlocality. In the bipartite case not all pure states, e.g., all the maximally entangled states, can exhibit  Hardy-type nonlocality while all tripartite entangled pure states exhibit Hardy-type nonlocality.

Second, for an arbitrary tripartite entangled pure state with the underlying Hilbert space having arbitrary dimensions, we have expansion
\begin{eqnarray}\label{pd}
|\psi_d\rangle=h_d|000\rangle+\sum_{j,k,l\not=0}t_{jkl}|jkl\rangle\hskip 2cm\nonumber\\
+\sum_{j,k\not=0}\big(u_{jk}|0jk\rangle+v_{jk}|j0k\rangle+s_{jk}|jk0\rangle\big).
\end{eqnarray}
in its magic basis with $h_d\not=0$.  If there exist $j,k,l\not=0$ such that $t_{jkl}\not=0$ then we make a local projection to the 3-qubit subspace spanned by $\{|0\rangle_1, |j\rangle_1\}\otimes\{|0\rangle_2,|k\rangle_2\}\otimes\{| 0\rangle_3,|l\rangle_3\}$.  If $t_{jkl}=0$ for all possible $j,k,l$ then, since $|\psi\rangle$ is fully entangled, we can suppose $u_{jk}\not=0$ and $s_{pq}\not=0$ for some $j,k,p,q$  without lose of generosity. After locally projecting qudit 1 and 2 to the 2-qubit subspace spanned by $\{|0\rangle,|p\rangle_1\}\otimes\{|0\rangle_2,|j\rangle_2\}$  we obtain
\begin{equation}\label{p3}
|\psi^\prime_d\rangle=h_d|000\rangle+|0j\phi\rangle+|p0\phi^\prime\rangle+ c_{pj}|pj0\rangle
\end{equation}
where $|\phi\rangle_3$ and $|\phi^\prime\rangle_3$ are two single-qudit states orthogonal to $|0\rangle_3$ with nonzero norms. Let a normalized $|\phi_+\rangle\propto|\phi\rangle$ if $|\phi\rangle$ is proportional to $|\phi^\prime\rangle$ and $|\phi_+\rangle_3\propto |\phi\rangle_3+|\phi^\prime\rangle_3$ if otherwise.  After a further local projection to the 2 dimensional subspace of the third qudit spanned by $\{|0\rangle_3,|\phi_+\rangle_3\}$, the first three terms in Eq.(\ref{p3}) are left, since $|\phi_+\rangle$ has non vanishing overlaps with both $|\phi\rangle$ and $|\phi^\prime\rangle$. In either cases we obtain a fully entangled 3-qubit state in its magic basis (may be different from the one defined by the closest product state) by a local projection: in the first case we have $t\not=0$ while in the second case we have $t=0$ with at least two coefficients among $u,v,s$ being nonzero. Therefore, according to the above discussion for 3-qubit pure states, we conclude  that every tripartite entangled pure state passes a Hardy-type test and therefore exhibits genuine tripartite nonlocality.

Third, our 3-particle Hardy-type test has a direct generalization to arbitrary number of particles. The original Hardy's test for two particles \cite{hardy} contains conditions  $P(aa)>0$ and $P(h)=0$ for every $h\in H_2=\{a\bar b, \bar b a,bb\}$ or simply $P(H_2)=0$. Our 3-particle Hardy-type test  can be written as $P(a^3)>0$ and $P(H_3)=0$ with $H_3=\{aH_2,bab,\bar b \bar b b\}$. Suppose that $P(a^{n-1})>0$ and $P(H_{n-1})=0$ are Hardy-type test for $n-1$ particles, then $P(a^n)>0$ and $P(H_n)=0$ where
\begin{eqnarray}
H_n&=&\{aH_{n-1},ba^{n-3}ab,\bar b a^{n-3} \bar b b\}
\end{eqnarray}
give rise to a Hardy-type test for $n$ particles which can be proved to be failed by all local non-signaling models  (see Appendix).

As a final remark, the maximal surviving probability $q_n=\max P(a^n)$ can be found analytically for arbitrary $n$ \cite{yu2}. In the case of three qubits we have
\begin{equation}
q_3= 
\frac{(1-\xi^2)\xi^2}{(2+\xi)^2}\approx0.0347513,
\end{equation}
with $\xi$ being the unique positive root of $x^3+4x^2-2=0$. We note that this bound is device independent. For another example in the following 3-qubit state
\begin{eqnarray}
|00+\rangle+|010\rangle-|10+\rangle-3|011\rangle-3|11+\rangle
\end{eqnarray}
if we choose  the same measurement settings $|a\rangle=|0\rangle$ and $|b\rangle\propto|+\rangle=|0\rangle+|1\rangle$, corresponding to the measurements of observables $\sigma_z$ and $\sigma_x$, respectively, for all three qubits then all conditions in Eq.(\ref{ht}) are satisfied with a near-optimal survival probability $1/32\approx0.03125$.

To conclude, with the help of a Hardy-type argument we have proved that tripartite entangled pure states do not admit a non-signaling local model and thus exhibit genuine tripartite nonlocality. In other words, in order to reproduce the correlations in a tripartite entangled pure state by a hybrid local/nonlocal model, at least one observer must have access to signaling correlations with others. Thus we have established a strong equivalence between genuine multipartite entanglement and nonlocality for tripartite pure states. It is highly possible to generalize this equivalence to  more than 3 particles.  In view of existing experimental implementations of Hardy's two-particle test \cite{dik,weak}, our 3-particle gedanken experiment is also within the reach of current technologies.

This work is supported by National Research Foundation and Ministry of Education, Singapore (Grant No. WBS: R-710-000-008-271).

\renewcommand{\theequation}{A.\arabic{equation}}
\setcounter {equation}{0}
\section{Appendix}

{\it Magic basis for a tripartite pure state ---} For a given 3-qudit pure state,
 a magic basis is a computational basis $\{|0\rangle,|1\rangle,\ldots\}\otimes \{|0\rangle,|1\rangle,\ldots\}\otimes \{|0\rangle,|1\rangle,\ldots\}$ under which
\begin{equation}\label{mb}
\langle \psi|000\rangle\not =0,\quad \langle\psi|\phi_k0_{k+1}0_{k+2}\rangle=0,
\end{equation}
for any pure state $|\phi\rangle_k$ that is orthogonal to $|0\rangle_k $ for $ k=1,2,3$ with indices $4,5$ being identified with 1,2, respectively. A magic basis of $|\psi\rangle$ always exists and, e.g., can be defined by its closest product state $|p\rangle=|p_1p_2p_3\rangle$ that maximizes $|\langle\psi|p\rangle|^2$ among all product states.
First, we have $h=\langle\psi|p\rangle\not=0$ since $|\langle\psi|p\rangle|^2$ is the largest overlap between $|\psi\rangle$ and a product state. Second, if there were some $k$, e.g., $k=3$, such that $g=\langle\psi|p_1p_2\phi_3\rangle\not=0$ for some  $|\phi\rangle_3$ orthogonal to $|p_3\rangle_3$, then we would have $|\langle\psi|p_{1}p_2\phi^\prime_3 \rangle|^2=|h|^2+|g|^2>|h|^2$ where $|\phi^\prime\rangle_3\propto h^*|p_3\rangle_3+g^*|\phi\rangle_3$ is normalized. This contradicts the definition of the closest product state so that conditions in Eq.(\ref{mb}) are satisfied, i.e., the computational basis with $|p_1p_2p_3\rangle$ taken as $|000\rangle$ is a magic basis.

{\it Proof of Theorem --- } For a given pure 3-qubit state, all possible  measurement settings satisfying the last four conditions in Eq(\ref{ht}) are those given in Eq.(\ref{ms}) with arbitrary complex $z$ and $(x,y)$.
In fact, by assuming $|a_1\rangle$ and $|b_3\rangle$ as in Eq.(\ref{ms}), it follows from condition Eq.(\ref{ht2}) that
$$0=\langle\psi|a_1b_2b_3\rangle=\langle b_2^*|C|a_1\rangle,$$
which leads to $\langle b_2^*|\propto\langle a_1^*|C^TJ$ since $\langle a_1^*|G^TJG|a_1\rangle=0$ for an arbitrary $2\times 2$ matrix $G$. From condition
$$0=\langle\psi|\bar b_1\bar b_2b_3\rangle=\langle\bar b_2^*|C|\bar b_1\rangle$$
it follows $|\bar b_1\rangle\propto JC^T|\bar b_2\rangle\propto JC^TC^*|a_1^*\rangle$, using $|b_2\rangle$ obtained above. From condition
$$0=\langle\psi|b_1a_2b_3\rangle=\langle a_2^*|C|b_1\rangle$$
it follows $\langle a_2^*|\propto \langle b_1^*|C^TJ\propto\langle a_1^*|C^TC^*C^TJ$, using $|b_1\rangle$ obtained above. From condition
$$0=\langle\psi|a_1\bar b_2 a_3\rangle=\langle\bar b_2^*|D|a_3\rangle$$ it follows $|a_3\rangle\propto JD^T|\bar b_2\rangle\propto JD^TC^*|a_1^*\rangle$.

All six states in Eq.(\ref{ms}) have nonzero norms for arbitrary $|a_1\rangle$ if $z$ makes $C$ invertible, which is obvious  except possibly for $|a_3\rangle$. Since $C|a_1\rangle=D|b_3\rangle$, we have $\langle \bar a_3|b_3\rangle=\langle a_1|C^\dagger C|a_1\rangle>0$ so that the state $|a_3\rangle$ also has a nonzero norm if $C$ is invertible. Recalling that the single qubit operators $C,D$ have matrix forms
$$C=\left(\begin{array}{cc}h&vz\\ uz& s+tz\end{array}\right),\quad D=\left(\begin{array}{cc}hx&vy\\ sy& ux+ty\end{array}\right)
.$$
Matrix $C$ is not invertible if and only if $$\det C=h(s+tz)-uvz^2=0,$$ which has at most two solutions of $z$ for a fully entangled $|\psi\rangle$ since $h\not=0$ and $t+s+uv>0$. If we choose $z$ such that $\det C\not=0$, as we will always do in what follows, at most 2 values of $z$ are excluded.

In a given pure state $|\psi\rangle$, each joint probability appears in Eq.(\ref{ht}), e.g., $P( a_1\bar b_2a_3)$, becomes proportional to, e.g., $|\langle\psi|a_1\bar b_2a_3\rangle|^2$ and so on. The second condition Eq.(\ref{ht4}) becomes Eq.(\ref{y}) which turns out to be a homogenous quadratic equation
\begin{equation}
\label{y1}x^2F_{00}+xy(F_{01}+F_{10})+y^2F_{11}=0
\end{equation}
in $x$ and $y$ where $F=C^TC^*C^TJ\tilde C$ with its matrix elements $F_{\mu\nu}=\langle\mu|F|\nu\rangle$ $(\mu,\nu=0,1)$ being independent of $x,y$. This is a quadratic equation of $x/y$ or $y/x$ if $F_{00}\not=0$ or $F_{11}\not=0$, respectively, with at leas one solution and has a solution  $x=0$ or $y=0$ if otherwise. That is, Eq.(\ref{y}) has at least one solution $(x,y)$ for any given $z$.

At this stage, by choosing the measurement settings as in Eq.(\ref{ms}) with $z$ not taking at most two values that makes $\det C=0$ and $(x,y)$ determined by Eq.(\ref{y}) for the given $z$, all conditions in Eq.(\ref{ht})  are satisfied except possibly for  the first condition, which reads
\begin{eqnarray}\label{eee}
|\langle\psi| a_1a_2 a_3\rangle|^2=|\langle a_1^*|C^TC^*C^TJDJD^TC^*|a_1^*\rangle|^2\nonumber\\
=\langle a_1^*|(C^TC^*)^2|a_1^*\rangle^2|\det D|^2>0.\hskip .1cm
\end{eqnarray}
Since we have chosen $z$ such that $\det C\not=0$ the above condition is equivalent to $\det D\not=0$. We have to avoid such values of $z$ that $(x,y)$ determined by Eq.(\ref{y}) also makes $\det D=0$.

Let us examine under what conditions the equation Eq.(\ref{y}) holds simultaneous with
$$\det D=hux^2+htxy-sv y^2=0,$$
which has at most two solutions $(x_\pm,y_\pm)$ for a fully entangled $|\psi\rangle$.  Denote by $D_\pm=D|_{(x,y)=(x_\pm,y_\pm)}$ and it holds $D_\pm^T=|p\rangle\langle q|$ with $|p\rangle$ and $|q\rangle$ having nonzero norms because $D_\pm$ is of rank 1 and is nonzero. By plugging $|a_{1\pm}\rangle=x_\pm|0\rangle+y_\pm|1\rangle$ into Eq.(\ref{y}) we obtain
\begin{eqnarray}\label{x}
&&\langle a^*_{1\pm}|C^TC^* C^TJ\tilde C|a_{1\pm}\rangle
\nonumber\\
&=&\langle b_3^*|D^T_\pm C^*\big(|a^*_{1\pm}\rangle\langle a^*_{1\pm}|+|\bar a^*_{1\pm}\rangle\langle \bar a^*_{1\pm}|\big) C^TJD_\pm|\bar b_3\rangle\nonumber\\
&=&\langle b_3^*|p\rangle\langle q|\tilde D_\pm^*|b_3^*\rangle\langle b_3^*|\tilde D_\pm^T|\bar q\rangle\langle p^*|\bar b_3\rangle=0
\end{eqnarray}
where
$$\tilde D_\pm=y^*D_0-x^*D_1=\left(\begin{array}{cc}hy^*&-vx^*\\ -sx^*& uy^*-tx^*\end{array}\right).$$ In the first equality above we have used the identities $C|a_{1\pm}\rangle=D_\pm|b_3\rangle$ and $\tilde C|a_{1\pm}\rangle=D_\pm|\bar b_3\rangle$. In the second equality we have used the identities $\langle\bar a^*_{1\pm}|C^T=\langle b_3^*|\tilde D^T_\pm$, $\langle a_{1\pm}^*|C^T=\langle b_3^*|D^T_\pm$, and $D^T_\pm JD_\pm=J\det D_\pm=0$.

It turns out that $\tilde D_\pm^T|\bar q\rangle$ has a nonzero norm because $\tilde D^T_\pm|\bar q\rangle\langle p^*|= \tilde D^T_\pm JD_\pm\not=0 $ which follows from the fact
$$\langle b_3^*|\tilde D^T_\pm JD_\pm|b_3\rangle=\langle \bar a_{1\pm}^*|C^TJC|a_{1\pm}\rangle=\det C\not=0.$$
And $\langle q|\tilde D_\pm^*$ has a zero norm if and only if $D^T_\pm\tilde D^*_\pm=0$ or
\begin{eqnarray*}
(|h|^2-s^2)x_\pm y_\pm&=&0,\\
-hvx^2_\pm+suy^2_\pm-stx_\pm y_\pm&=&0,\\
 h^*v y^2_\pm-sux^2_\pm-tsx_\pm y_\pm&=&0,\\
 (u^2-t^2-v^2)x_\pm y_\pm+ut(y^2_\pm-x^2_\pm)&=&0.
 \end{eqnarray*}
The first equation holds if and only if either  $x_\pm y_\pm=0$ or $x_\pm y_\pm\not=0$  and $|h|=s$. If $x_\pm y_\pm=0$ then we have $v=su=ut=0$ from the last three equations, meaning that the state has to be symmetric with $s=u=v=0$, which corresponds to GHZ-like states. If $|h|=s$ then the state has to be symmetric with $x_\pm,y_\pm$ satisfying $s^2y_\pm^2=hx_\pm(sx_\pm+ty_\pm)$. From the second equation above we obtain $(h-s)t=0$, from which it follows either i) $h=s$ or ii) $t=0$ leading to $|h|=s$ due to the third equation above. In these three cases the solutions to Eq.(\ref{y}) coincide with $(x_\pm,y_\pm)$ so that our test fails. Otherwise, especially for asymmetric states, Eq.(\ref{x}) has exactly four solutions of $z$ for each $(x_\pm,y_\pm)$. Thus, by excluding at most 8 values of $z$, equations Eq.(\ref{y}) and Eq.(\ref{eee}) hold simultaneously, i.e., the test is passed.

To summarize, for asymmetric pure states and those symmetric states specified in the first part of Theorem, we have only to exclude at most 2+8=10 values of $z$ to satisfy both Eq.(\ref{y}) and Eq.(\ref{eee}), i.e., all conditions in Eq.(\ref{ht}) are satisfied.  \hfill $\square$

{\it A Hardy-type test for symmetric 3-qubit states  ---} The Hardy-type test \cite{chenq} for three particles  includes six conditions in which five of them coincide with the first five conditions in Eq.(\ref{ht}) and the other condition is $P(\bar b_1a_2a_3)=0$. For every bipartite cut there is always a set of four conditions that is a conditioned Hardy's 2-particle test and thus all non-signaling local models fail this Hardy-type test. For an arbitrary symmetric pure state
$$|\psi_s\rangle=h^*|000\rangle+s(|011\rangle+|101\rangle+|110\rangle)+t|111\rangle,$$
if we choose the measurement settings for the first two qubits to be identical, then
only the first four conditions in Eq.(\ref{ht}) has to be satisfied. Consider the measurement settings $|a_1\rangle=|a_2\rangle=|x\rangle:=|0\rangle+x|1\rangle$ and $|b_1\rangle=|b_2\rangle=JDD^\dagger|x^*\rangle$ together with  $|b_3^*\rangle= D^T|x\rangle$ and $|a_3\rangle=JD^TD^*D^T|x\rangle$ with $x$ being arbitrary for the moment. It can readily be checked that all three conditions Eq.(\ref{ht4}-\ref{ht2}) are satisfied. As a result
\begin{eqnarray*}
\langle\psi_s|a_1a_2a_3\rangle&=&\langle x^*|DJD^TD^*D^T|x\rangle\\
&=&\langle\bar x|D^*D^T|x\rangle\det D:= R(x)\det D.
\end{eqnarray*}
Here $R(x)$ is a polynomial of a degree $\le 4$ for $|x|$ with the phase $e^{i\gamma}=x/|x|$  fixed. The constant term of $R(x)$ vanishes while the linear term of $R(x)$ reads $(|h|^2-2s^2)x-hsx^*$ which is nonzero if we take $2\gamma\not=\beta,\beta+\pi$ with $e^{i\beta}=h/|h|$.  Thus $R(|x|e^{i\gamma})=0$ has at most 4 roots for $|x|$. For entangled $|\psi_s\rangle$ the equation $\det D=0$ is a quadratic equation for $x$ with at most two solutions for $x$. If we take $\gamma$ as specified above and exclude at most 4+2=6 values of $|x|$ then we can always ensure $\langle\psi_s|a_1a_2a_3\rangle\not=0$ so that $P(a_1a_2a_3)>0$, i.e., the first condition is also satisfied.

{\it Hardy-type test for $n$ particles ---}
Consider a system composed of $n$  subsystems possessed by $n$ distant observer labeled with the index set $I=\{1,2,\ldots,n\}$. Each observer $k\in I$ measures two dichotomic observables $\{a_k,b_k\}$ with outcome labeled with $\{0,1\}$. Let ${\mathcal P}_I$ denote the set of all nontrivial partitions of $I$, i.e., unordered pairs $\{\alpha,\beta\}$ such that $\alpha\cup\beta =I$ and $\alpha\cup\beta=\emptyset$ with $\alpha,\beta\subset I$. In the most general hybrid local/nonlocal model, the joint probability $P(a_I)$ of observer $k$ measuring observable $a_k$ with outcome $0$ for all observers $k\in I$ assumes the following bi-local form
\begin{equation}\label{bln}
P(a_I)=\sum_{\{\alpha,\beta\}\in{\mathcal P}_I}\int d\lambda\ \varrho_\lambda^\alpha L_\lambda^\alpha(a_\alpha)S_\lambda^\beta(a_\beta).
\end{equation}
Here $S_\lambda^{\alpha}$ and $L_\lambda^{\beta}$ are some multipartite probability distributions on subsystems $\alpha$ and $\beta$, respectively,  for a given hidden variable $\lambda$  distributed according to $\varrho_\lambda^{\alpha}$, respectively. Those multipartite correlations are further restricted by non-signaling conditions, e.g., they should satisfy conditions similar to Eq.(\ref{ns}).

Hardy's test for two particles can be simply denoted by  $P(aa)>0$ and $P(h)=0$ for all  $h\in H_2=\{a\bar b, \bar b a,bb\}$ or simply $P(H_2)=0$. And our 3-particle Hardy-type test can be written as $P(aaa)>0$ and $P(H_3)=0$ with $H_3=\{aH_2,bab,\bar b \bar b b\}$. Suppose that $P(a^{n-1})>0$ and $P(H_{n-1})=0$ are Hardy-type test for $n-1$ particles, then $P(a^n)>0$ and $P(H_n)=0$ where
\begin{eqnarray*}
H_n&=&\{aH_{n-1},ba^{n-3}ab,\bar b a^{n-3} \bar b b\}\\
&=&\{a_Ja\bar b, a_J\bar ba,a_Jbb, a_{J\setminus k}b_kab,a_{J\setminus k}\bar b_k\bar b b\}_{k\in J}
\end{eqnarray*}
with $J=\{1,2,\ldots,n-2\}$,  are the Hardy-type test for $n$ particles, i.e., these $2n$ conditions are incompatible with the bi-local form Eq.(\ref{bln}). In fact,
being a Hardy-type test for last $n-1$ particles conditioned on the first observer's measuring observable $a$ with $0$ as outcome, the first $2(n-1)$ conditions $P(a^n)>0$ and $P(aH_{n-1})=0$ cannot be satisfied if the last $n-1$ particles are not fully correlated. This leaves us a unique possibility that the first particle is isolated while all the other $n-1$ particles may share some multipartite correlations. From the condition $P(a^n)>0$ it follows that there exist some hidden variables $\lambda$ such that $L_\lambda(a_1)S_\lambda(a_2\ldots a_n)>0$. This is impossible because of the following contradiction
\begin{eqnarray}
0&=&L_\lambda(\bar b)S_\lambda(a^{n-3}\bar bb)+L_\lambda(a)S_\lambda(a^{n-3}bb)\nonumber\\
&\ge&\min\{L_\lambda(\bar b),L_\lambda(a)\}[S_\lambda(a^{n-3}\bar bb)+S_\lambda(a^{n-3}bb)]\nonumber\\
&=&\min\{L_\lambda(\bar b),L_\lambda(a)\}[S_\lambda(a^{n-3}\bar ab)+S_\lambda(a^{n-3}ab)]\nonumber\\
&\ge&[L_\lambda(a)-L_\lambda(b)]S_\lambda(a^{n-3}ab)\nonumber\\
&=&L_\lambda(a)[S_\lambda(a^{n-3}ab)+S_\lambda(a|a^{n-3}a\bar b)]\nonumber\\
&=&L_\lambda(a)[S_\lambda(a^{n-3}aa)+S_\lambda(a|a^{n-3}a\bar a)]\nonumber\\
&\ge&L_\lambda(a)S_\lambda(a^{n-2})>0.
\end{eqnarray}
Here the first equality is because both $\bar ba^{n-3}\bar bb$ and $a^{n-2}bb$ belong to $H_n$ and the third equality equality is because $ba^{n-3}ab$ and $a^{n-1}b$ belong to $H_n$ while the second and the fourth equalities are due to non-signaling conditions.  Thus we conclude that the bi-local form Eq.(\ref{bln}) is incompatible with our $n$-particle Hardy-type test.

\end{document}